# Synthetic Observations of the Infalling Rotating Envelope: Links between the Physical Structure and Observational Features

Shoji Mori[1,2], Yuri Aikawa[2], Yoko Oya[3], Satoshi Yamamoto[4,5], and Nami Sakai[6]  
[1] Astronomical Institute, Graduate School of Science, Tohoku University, 6-3 Aoba, Aramaki, Aoba-ku, Sendai, Miyagi, 980-8578, Japan; mori.s@astr.tohoku.ac.jp  
[2] Department of Astronomy, Graduate School of Science, The University of Tokyo, 113-0033, Tokyo, Japan  
[3] Center for Gravitational Physics and Quantum Information, Yukawa Institute for Theoretical Physics, Kyoto University, Kyoto, 606-8502, Japan  
[4] The Graduate University for Advanced Studies SOKENDAI, Shonan Village, Hayama, Kanagawa 240-0193, Japan  
[5] Research Center for the Early Universe, The University of Tokyo, 7-3-1, Hongo, Bunkyoku, Tokyo 113-0033, Japan  
[6] RIKEN Cluster for Pioneering Research, 2-1, Hirosawa, Wako-shi, Saitama 351-0198, Japan  
Received 2022 October 30; revised 2023 September 28; accepted 2023 October 2; published 2024 January 11

## Abstract

We performed synthetic observations of the Ulrich, Cassen, and Moosman (UCM) model to understand the relation between the physical structures of the infalling envelope around a protostar and their observational features in molecular lines, adopting L1527 as an example. We also compared the physical structure and synthetic position–velocity (P–V) diagrams of the UCM model and a simple ballistic (SB) model. There are multiple ways to compare synthetic data with observational data. We first calculated the correlation coefficient. The UCM model and the SB model show similarly good correlation with the observational data. While the correlation reflects the overall similarity between the cube datasets, we can alternatively compare specific local features, such as the centrifugal barrier in the SB model or the centrifugal radius in the UCM model. We evaluated systematic uncertainties in these methods. In the case of L1527, the stellar mass values estimated using these methods are all lower than the value derived from previous Keplerian analysis of the disk. This may indicate that the gas infall motion in the envelope is retarded by, e.g., magnetic fields. We also showed analytically that, in the UCM model, the spin-up feature of the P–V diagram is due to the infall velocity rather than the rotation. The line-of-sight velocity $V$ is thus $\propto x^{-0.5}$, where $x$ is the offset. If the infall is retarded, rotational velocity should dominate so that $V$ is proportional to $x^{-1}$, as is often observed in the protostellar envelope.

*Unified Astronomy Thesaurus concepts:* Star formation (1569); Protostars (1302); Astrochemistry (75)

## 1. Introduction

The Atacama Large Millimeter/submillimeter Array (ALMA) is revealing circumstellar structures of low-mass protostars at spatial scales of a few tens of au in nearby star-forming regions (e.g., Ohashi et al. 2014, 2023; Sakai et al. 2014b; Oya et al. 2016; Aso et al. 2017; Yen et al. 2017; Okoda et al. 2018, 2021; Jacobsen et al. 2019; Tobin et al. 2019; Bianchi et al. 2020; Harsono et al. 2021). In the vicinity of a protostar, a rotationally supported disk is formed from infalling material with angular momentum. While density distributions can be estimated from dust continuum observations, molecular line observations are essential to reveal the gas dynamics, such as the infall velocity of the envelope gas, the spin-up of the infalling material, and disk formation.

While CO is stable and abundant in a large fraction of the volume around a protostar (e.g., Aso et al. 2017; van't Hoff et al. 2020), lines of other molecules trace specific regions/structures (e.g., Oya et al. 2016, 2017; Garufi et al. 2020; Pineda et al. 2020; Tychoniec et al. 2021; Lin et al. 2023; Yamato et al. 2023). SO is found in a transitional ring region between the envelope and Keplerian disk, e.g., around the class 0/I protostar L1527 (Sakai et al. 2014b, 2017; van't Hoff et al. 2023) and the class I protostar IRAS 04365 + 2535 in TMC-1A (Sakai et al. 2016; Tychoniec et al. 2021), possibly tracing a weak accretion shock with respect to the radial velocity (Aota et al. 2015; Miura et al. 2017; van Gelder et al. 2021; Shariff et al. 2022).

Unsaturated hydrocarbons such as $C_2H$ and $c-C_3H_2$, on the other hand, are abundant in the infalling rotating envelope but sharply decrease at or around the transition region traced by SO (Sakai et al. 2014b, 2014b; Okoda et al. 2018). $C_2H$, for example, reacts with $H_2$ to convert to $C_2H_2$; the activation barrier can be overcome by shock heating. Hereafter, we refer to such molecules that are destroyed in disk regions as envelope tracing molecules (ETMs). These spatial variations of molecular abundances are found not only in integrated intensity maps, but also in position–velocity diagrams (P–V diagrams hereafter). The kinematic structures traced by the CO, SO, and $C_2H$ lines in L1527 are quite different from each other (Sakai et al. 2014b). This means that we can trace specific regions (e.g., disk and envelope) by choosing appropriate molecular lines.

P–V diagrams are frequently used to analyze the kinematics and basic physical properties of protostellar systems. One of the basic parameters for star formation is the mass of the central star. For example, Aso et al. (2017) analyzed the P–V diagram of the $C^{18}O$ ($J = 2 - 1$) line toward L1527 to show that the rotation velocity inside a radius of $\sim 74$ au is proportional to $\sim R^{-0.5}$, where $R$ is the cylindrical radius measured from the central star. They derived the dynamical mass of the central star to be $\sim 0.45\,M_\odot$ from the P–V diagram by assuming Keplerian rotation. This is slightly larger than the value $0.3\,M_\odot$ estimated by Ohashi et al. (2014) with a lower spatial resolution. Ohashi et al. (2014) reported that the midplane infall velocity in the envelope has to be smaller than the free-fall velocity by a factor of ∼2 outside 250 au in order to reproduce the characteristic features of the observed line profiles and P–V diagrams (see Figures 1 and 2 in Ohashi et al. 2014). Reduced infall velocity of the envelope gas is similarly suggested around L1551 IRS5 (Takakuwa et al. 2013; Chou et al. 2014) and TMC-1A (Aso et al. 2015).

Sakai et al. (2014b), on the other hand, analyzed the P–V diagram of unsaturated hydrocarbons, which selectively trace the envelope in L1527. The P–V diagram of c-$C_3H_2$ emission (Sakai et al. 2014b) shows a spin-up of angular velocity from the offset of 700–100 au. This emission disappears at ∼100 au, which is interpreted as a centrifugal barrier, i.e., the perihelion of a ballistic particle of the infalling rotating envelope, which is used to derive the stellar mass to be 0.18 $M_\odot$. In this model, the envelope gas particles infall with a ballistic orbit and stop infalling at the barrier. We shall refer to this as the simple ballistic (SB) model (see Section 3.1 and Oya et al. (2022) for a more detailed description). Similar analyses have been performed for several objects in recent years (Oya et al. 2014, 2015; Sakai et al. 2016; Zhang et al. 2018).

Both of these methods of estimating stellar mass have pros and cons. While the estimate based on Keplerian rotation (rotation velocity $\propto R^{-1/2}$) seems simple and robust, CO is abundant both in the disk and the envelope. As a result, the P–V diagram is a mixture of both velocity components. It has been discussed how we can best derive the Keplerian rotation (e.g., Ohashi et al. 2014; Aso et al. 2017; Aso & Machida 2020; Maret et al. 2020; Sai et al. 2022, 2023). High angular resolution observations are necessary to resolve the velocity profile in the disk region. In addition, in the cold and dense region of the disk, freezing out of CO changes the abundance of CO, complicating the derivation of the circumstellar structure (e.g., Tychoniec et al. 2021). Hydrocarbons, on the other hand, are expected to selectively trace the envelope. The analysis of gas velocities of the envelope is especially important when the disks are too small to be spatially resolved or deeply embedded.

While the SB model has been successfully applied to several protostellar cores, the velocity structure of infalling rotating envelopes is not as simple as Keplerian rotation, and the derived stellar mass could depend on the assumed envelope model. For example, one of the arguments against the SB model is that a radial shock (i.e., a shock that slows the radial velocity to subsonic speeds) could occur before the centrifugal radius is reached, thus rendering the ballistic approximation invalid (Shariff et al. 2022). Also, the infalling envelope around the disk could be decelerated by the gas pressure and viscosity (Jones et al. 2022). A shock around or outside the centrifugal radius is observed in (magneto-)hydrodynamic simulations (Machida et al. 2010; Zhao et al. 2016; Jones et al. 2022) and is suggested by observations (e.g., Ohashi et al. 2014; Aso et al. 2017; Sakai et al. 2017). If the shocked region (where hydrocarbons can be destroyed) is interpreted as a centrifugal barrier assuming the SB model, the mass of the central protostar will be underestimated. Also, the SB model often assumes that the brightness distribution of the envelope is given by a power law of the distance from the central star for simplicity (e.g., Oya et al. 2014, 2015, 2022; Okoda et al. 2018). It could be a good approximation in the very early stage of the core, e.g., similar to the singular isothermal sphere (Shu 1977), but the density distribution deviates from a single power law as the collapse proceeds. The gas temperature could also deviate from a power law; for instance, if the outflow cavity wall is preferentially heated by the irradiation from the central star.

In the present work, we perform synthetic observations of the standard hydrodynamical model of an infalling rotating envelope by Ulrich (1976) and Cassen & Moosman (1981) (the UCM model; see also Terebey et al. 1984). Our objectives are twofold. First, we investigate the observational features obtained from the UCM model and the correspondence between the physical structure and the observational features of the envelope. Second, we investigate the uncertainty of the stellar mass estimate due to the choice of envelope models using the SB and UCM models applied to L1527 as an example.

We perform radiative transfer calculations to obtain a temperature structure consistent with the density structure in the protostellar envelope (e.g., Agurto-Gangas et al. 2019; Flores-Rivera et al. 2021), and to derive the cube data (channel map) of the emission line of ETMs. We analyze the cube data and the P–V diagram to see how they depend on model parameters. The models are then compared with the observational data of L1527 in the c-$C_3H_2$ emission, which preferentially traces the envelope gas, to derive the mass of the central star and the angular momentum of the infalling gas. A similar comparison is made with the SB model to see how the derived stellar mass depends on the assumed envelope model.

This paper is organized as follows. In Section 2, we describe the basic assumptions and settings of the hydrodynamics model and radiation transfer calculations. In Section 3, we describe our fiducial model and the major properties of its P–V diagram. Furthermore, we quantitatively compare the P–V diagrams of the synthetic observations with those obtained from the observation of L1527 (Sakai et al. 2014b) to derive the best-fit mass of the central star. In Section 4, we discuss the velocity profile, the mass estimation focusing on the specific features in the P–V diagram, and the physical structures of the envelope around L1527. In Section 5, we summarize the paper.

## 2. Methods

We perform synthetic observations of an infalling rotating envelope around a protostar. First, we calculate the density and velocity structures in the envelope following the UCM model. Temperature structure is determined by solving the radiation transfer. We then produce the cube data of an emission line from ETMs by calculating radiation transfer using a ray-tracing method, in which the radiative intensity is integrated along a line of sight. We analyze the cube data to derive the integrated intensity map and P–V diagram. We have developed an envelope observation simulator, envos (Mori 2023), which is a Python tool that simplifies the above procedures, from constructing physical models to conducting synthetic observations.

### 2.1. Calculation of Kinematic Structures

We adopt the classical infalling envelope model (the UCM model; Ulrich 1976; Cassen & Moosman 1981; see also Terebey et al. 1984). The UCM model provides the density and velocity distribution of the infalling envelope gas at a given time in a quasi-steady state, based on the following assumptions.

1. The molecular cloud core before the gravitational collapse is a singular isothermal sphere with rigid rotation. The gravitational collapse starts from the center and the collapsing surface propagates outward at the sound speed $c_{\rm cl}$ (Shu 1977), depending on the cloud temperature $T_{\rm cl}$. The gas temperature during the collapse is also assumed to be isothermal.
2. The envelope structure is axisymmetric about the rotation axis and plane-symmetric with respect to the equatorial plane, which is perpendicular to the rotation axis.
3. The gas particles travel in a ballistic orbit around the protostar, conserving their angular momentum and total mechanical energy because gravity is the dominant force on the infalling gas until it reaches the disk, which lies in the equatorial plane.

We assume that the abundance of ETMs sharply drops when the fluid parcel reaches the disk, e.g., via shock heating (see Section 1), although the shocked region could be larger than the disk size (Shariff et al. 2022). We thus focus only on the physical structure of the envelope and do not describe the structure of the disk. We neglect the effect of the disk on the envelope structure by assuming that the Keplerian disk is infinitely thin. The UCM model assumes that the total stellar mass is the sum of the material that has reached the central star or the equatorial plane. The gravitational potential of the disk is neglected, as the disk's gravity would not significantly alter the velocity of the infalling gas.

Since the model is axisymmetric, the physical parameters depend only on the distance $r$ from the star and the angle $\theta$ from the rotation axis, which is the polar axis. The velocity distribution is described in the spherical coordinate system as

$$v_r(r, \theta) = -\left(\frac{GM}{r}\right)^{1/2}\left(1 + \frac{\cos\theta}{\cos\theta_0}\right)^{1/2}, \quad (1)$$

$$v_\theta(r, \theta) = \left(\frac{GM}{r}\right)^{1/2}\left(\frac{\cos\theta_0 - \cos\theta}{\sin\theta}\right)\left(1 + \frac{\cos\theta}{\cos\theta_0}\right)^{1/2}, \quad (2)$$

$$v_\phi(r, \theta) = \left(\frac{GM}{r}\right)^{1/2}\left(\frac{\sin\theta_0}{\sin\theta}\right)\left(1 - \frac{\cos\theta}{\cos\theta_0}\right)^{1/2}, \quad (3)$$

where $G$ is the gravitational constant, $M$ is the protostellar mass, and $\theta_0(r, \theta)$ is the polar angle at which the gas, now at position $(r, \theta)$, was before the cloud collapse (Ulrich 1976; Terebey et al. 1984). Thus, the lines for constant $\theta_0$ show the streamlines of the infalling envelope. The density is given as

$$\rho(r, \theta) = -\frac{\dot{M}}{4\pi r^2 v_r}[1 + \zeta(r)(2 - 3\sin^2\theta_0)]^{-1}, \quad (4)$$

where $\dot{M}$ is the mass accretion rate of the envelope gas at the collapsing surface, and $\zeta$ is the inverse of $r$ normalized by the centrifugal radius $R_{\rm CR}$, $\zeta = R_{\rm CR}/r$. The centrifugal radius $R_{\rm CR}$ is defined as the location where the gravitational force is equal to the centrifugal force in the equatorial plane. When the angular momentum at the equatorial plane ($\theta_0 = \pi/2$) is $j_{\rm mid}$, $R_{\rm CR}$ is given as

$$R_{\rm CR} = \frac{j_{\rm mid}^2}{GM}. \quad (5)$$

The envelope gas reaches the equatorial plane and concentrates at $r \leqslant R_{\rm CR}$. As the maximum angular momentum of the gas accreting onto the disk is $\sqrt{GMR_{\rm CR}}$, the centrifugal radius is expected to be close to the disk radius (Cassen & Moosman 1981; Stahler et al. 1994; Shariff et al. 2022).

The density and velocity distributions (Equations (1)–(4)) can be determined once the $\theta_0$ distribution is known, which can be obtained from the orbital equations of the gas particles:

$$\zeta(r) = \frac{\cos\theta_0 - \cos\theta}{\sin^2\theta_0\cos\theta_0}. \quad (6)$$

Alternatively, the equation can be written as a cubic equation for $\cos\theta_0$,

$$\zeta(r)\cos^3\theta_0 + (1 - \zeta(r))\cos\theta_0 - \cos\theta = 0, \quad (7)$$

which can be solved analytically. Eliminating the solutions of the flow in the equatorial plane and the flow passing through the equatorial plane, we obtain one unique solution from Equation (7). We note that the velocity distribution scales with the centrifugal radius $R_{\rm CR}$ and velocity $V_{\rm CR} = \sqrt{GM/R_{\rm CR}}$ at the radius. The orbital equation is a function of $r/R_{\rm CR}$ and $\theta$, and thus $\theta_0$ is also a function of them. Therefore, the velocity at $(r, \theta)$ normalized with $V_{\rm CR}$ depends only on $r/R_{\rm CR}$ and $\theta$.

The physical parameters $R_{\rm CR}$, $M$, and $\dot{M}$ are related to each other, according to Shu's self-similar solution (Shu 1977). The sound speed in the envelope $c_{\rm cl}$ is

$$c_{\rm cl} = \sqrt{\frac{k_B T_{\rm cl}}{\mu m_p}}, \quad (8)$$

where $T_{\rm cl}$ is the temperature of the cloud core, $k_B$ is the Boltzmann constant, $m_p$ is the mass of a proton, and the mean molecular weight $\mu$ is set to be 2.3. During the cloud collapse, the collapsing surface propagates inside out at the sound velocity $c_{\rm cl}$. Since a steady state is assumed, the mass accretion rate at the surface is equal to that inside the surface. Considering Shu's solution for the density $\rho$ and radial velocity $v_r$, the mass accretion rate depends only on $c_{\rm cl}$,

$$\dot{M} = \frac{c_{\rm cl}^3 m_0}{G}, \quad (9)$$

where $m_0$ is a constant equal to 0.975 (Shu 1977). Once $\dot{M}$ is given, the central mass $M$ is related to the time $t$ since the start of the cloud collapse as

$$M = \dot{M}t. \quad (10)$$

Considering the angular momentum conservation for the gas infalling from the collapsing surface at $t$, the angular momentum for a streamline $\theta_0$ is

$$j(\theta_0) = \Omega_{\rm cl}\left(\frac{1}{2}m_0 c_{\rm cl} t \sin\theta_0\right)^2, \quad (11)$$

where $\Omega_{\rm cl}$ is the angular velocity of the cloud core. Thus, $j_{\rm mid}$ is written as

$$j_{\rm mid} = \Omega_{\rm cl}\left(\frac{1}{2}m_0 c_{\rm cl} t\right)^2. \quad (12)$$

In summary, by choosing one parameter each from the following three parameter sets $\{T_{\rm cl}, c_{\rm cl}, \dot{M}\}$, $\{M, t\}$, and $\{\Omega_{\rm cl}, R_{\rm CR}\}$, we can calculate the density and velocity distributions in the envelope.

**Table 1**
Summary of Fiducial Parameters

| Parameter | Value | Description |
| --- | --- | --- |
| $M$ ($M_\odot$) | 0.2 | Protostellar mass |
| $R_{CR}$ (au) | 200 | Centrifugal radius |
| $\dot{M}$ ($M_\odot \mathrm{yr}^{-1}$) | $4.5 \times 10^{-6}$ | Mass accretion rate |
| $\theta_{cav}$ (°) | 45 | Cavity angle |
| $L$ ($L_\odot$) | 2.75 | Luminosity of protostar |
| $i$ (deg) | 95 | Inclination angle |
| $PA_{object}$ (°) | 95 | Position angle of minor axis of the object |
| $a_{beam}$ (au) | 109.9 | Beam major axis |
| $b_{beam}$ (au) | 94.8 | Beam minor axis |
| $PA_{beam}$ (°) | 179.3 | Beam position angle |
| $\Delta v_{conv}$ (km s$^{-1}$) | 0.147 | Velocity resolution |

While the UCM model is often used as a standard model of gravitational collapse with angular momentum conservation, some assumptions may be too simplistic. For example, it assumes the rigid rotation of a core in the initial condition, while the angular momentum distribution within a molecular cloud and a cloud core is still debated (e.g., Larson 1981; Goodman et al. 1993; Burkert & Bodenheimer 2000; Caselli et al. 2002; Misugi et al. 2019; Gaudel et al. 2020; Heimsoth et al. 2022; Sai et al. 2022). Also, the angular momentum might be significantly reduced in the infalling gas due to angular momentum transport by the magnetic field (Li & McKee 1996; Tassis & Mouschovias 2005; Kunz & Mouschovias 2010). These issues will be discussed in Section 4.4.

In order to mimic an outflow cavity, we remove the gas for $\theta_0 < \theta_{cav}$, with $\theta_{cav}$ being the cavity angle. Although the outflow and thus the cavity could be more collimated, we expect the cavity shape is reasonably mimicked in the central region of the core ($\lesssim 1000$ au), in which we are interested.

In our calculations, the radial direction is divided into 460 cells on a logarithmic scale from 10 to 1000 au, and the polar angle direction is divided into 314 cells from 0° to 180°. The resolution does not affect the final outputs as long as it is sufficiently high compared to the convolution width (i.e., beam size and velocity resolution of a synthetic observation). Specifically, a resolution of about 1 au at a radius of 100 au is sufficient.

The model parameters and our fiducial values are summarized in Table 1. We basically refer to the values derived from observations of L1527. The opening angle of the outflow cavity is set to be 45° based on the Spitzer image of Tobin et al. (2008; see also van't Hoff et al. 2023). The mass accretion rate, which determines the density distribution, is set to be $4.5 \times 10^{-6} M_\odot \mathrm{yr}^{-1}$, based on the best-fit value of Tobin et al. (2013), though the normalized intensity distribution of the P–V diagram should be independent of it (see Section 2.3). The cloud core temperature $T_{cl}$, which determines the mass accretion rate, is 20 K. The central star mass and centrifugal radius chosen here are close to the values of L1527 previously reported (Sakai et al. 2014b): $M$ is 0.2 $M_\odot$ and $R_{CR}$ is 200 au (see Section 3.4).

### 2.2. Calculation of Thermal Structure

We calculate the radiation transfer using the Monte Carlo method with RADMC-3D (Dullemond et al. 2012) to obtain the temperature structure for a given density structure (Section 2.1). The photon packages are generated at random wavelengths and directions from the central star, absorbed by the dust, and then randomly emitted again as dust thermal radiation. By repeating this process, the temperature distribution is calculated from the final energy balance of each cell. We set the photon package number to be $10^7$, with which we have confirmed that the temperature structure converges. We ignore the scattering because the absorption coefficient dominates over the scattering coefficient for the thermal emission, where the wavelength is $\sim 100~\mu$m ($T \sim 30$K), unless the maximum grain size is larger than $\sim 100~\mu$m (Draine & Lee 1984; Miyake & Nakagawa 1993). The gas temperature is assumed to be the same as the dust temperature.

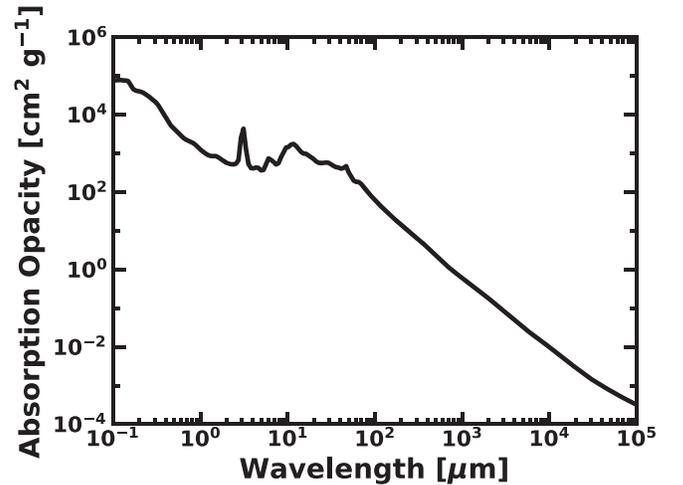

**Figure 1.** Dust absorption opacity as a function of wavelength used in the calculation of the thermal structure. The opacity is calculated with the code used in Birnstiel et al. (2018). Here, we assume the dust size distribution of Mathis et al. (1977).

In this study, we assume a dust-to-gas mass ratio of 0.01, i.e., a typical value of the interstellar medium. We calculate the absorption coefficient of dust grains using the code of Birnstiel et al. (2018) (Figure 1), assuming the grain size distribution in the interstellar medium (Mathis et al. 1977); the number density $n(a)$ of dust with grain size $a$ is $n(a) \propto a^{-3.5}$ for the range $0.005~\mu$m $< a < 0.25~\mu$m. Indeed, the opacity index of the dust in the outer disk region ($r \approx 50$–80 au) is suggested to be $\approx 1.67$ (Ohashi et al. 2022), which is the same as that of the interstellar medium. Therefore, the dust size of the envelope would be similar to that of the interstellar medium. The dust consists of water (20 wt%), silicon (32.9 wt%), troilite (7.4 wt%), and organic matter (39.7 wt%), following Birnstiel et al. (2018). The optical constants of Henning & Stognienko (1996), Draine (2003), and Warren & Brandt (2008) are adopted.

From the fiducial parameters (see Table 1), the luminosity of the central star is set to be 2.75 $L_\odot$ based on the observation of L1527 (Tobin et al. 2008). Note that, while the stellar luminosity controls the absolute value of the temperature, the normalized intensity distribution in the P–V diagram, which we focus on, is not strongly dependent on it.

### 2.3. Synthetic Observation

For the model described in Sections 2.1 and 2.2, we perform synthetic observations by calculating the radiation transfer for a ray-tracing method with RADMC-3D (Dullemond et al. 2012). For simplicity, local thermodynamic equilibrium (LTE) is assumed for the line emission. After calculating the radiation transfer with a fine resolution, we convolve the cube data with a Gaussian function in three dimensions (space and velocity directions) to account for the finite spatial and spectral resolutions of the observation by Sakai et al. (2014b) (see Table 1). Then, analyzing the cube data, we produce the integrated intensity map and the P–V diagram to be compared with those from the observation.

In L1527, the infalling rotating envelope is traced by the emission lines of hydrocarbons (Sakai et al. 2014b, 2014a). Among those lines, we chose to simulate the $5_{23}$–$4_{32}$ transition of c-$C_3H_2$, since it does not suffer the blending of other lines. The molecular line data are adopted from the Leiden Atomic and Molecular Database (Schöier et al. 2005). We ignore the dust continuum in the synthetic observation by assuming that the envelope is optically thin for the observation wavelength, and thus the dust absorption and scattering effects are ignored. The absolute value of the intensity depends on the assumed abundance of the ETMs. In Sakai et al. (2014b), the P–V diagram of the ETMs is point-symmetric with respect to the position of the protostar (see also Section 3.4), which suggests that their emission is mostly optically thin. In order to simplify the analysis, the molecular abundance is set to be low enough for the line to be optically thin. Furthermore, in the calculation of the P–V diagram, we normalize the diagram by the maximum brightness, so that we focus on the normalized intensity distribution of the diagram. More realistic simulation needs a detailed modeling of chemistry within the infalling envelope (e.g., Flores-Rivera et al. 2021), which is left for future work. We assume the inclination angle between the rotation axis and direction toward the observer to be 95°[7] (Oya et al. 2015), and the position angle of the major axis to be 5°.

We briefly present how the intensity at a pixel is calculated in the radiative transfer calculation. In this study, we assume no scattering and make use of the LTE approximation, resulting in the source function being equivalent to the Planck function $B_\nu$. Thus, the emissivity for a wavelength $\nu$ is written as $\kappa_\nu \rho B_\nu$, where $\kappa_\nu$ is the opacity. In this case, the radiative transfer equation under the optically thin limit is given by

$$\frac{dI_\nu}{ds} = \kappa_\nu \rho B_\nu. \quad (13)$$

For each pixel on the plane of the sky, the radiation intensity is integrated along the line of sight, giving the observational flux. If the Rayleigh–Jeans approximation is applicable (i.e., $B_\nu \approx 2\nu^2 k_B T/c^2$, where $c$ is the speed of light), the emissivity $\kappa_\nu \rho B_\nu$ is $\propto \rho T$, which is used for analysis in Section 3.2.

### 3. Results

The physical structure of the UCM model is first described in Section 3.1. We investigate how observational features depend on the physical structure (Section 3.2) as well as their dependence on the physical parameters and observational configurations (Section 3.3). In Section 3.4, we describe how well the simulated P–V diagram can explain the actual observations by comparing the model P–V diagram with that observed in L1527.

---

[7] In our model, the inclination is in the range of $0° \leqslant i \leqslant 180°$, since we define $i$ as the inclination of the angular momentum vector.

### 3.1. Physical Structures of the Infalling Envelope Models

In this section, we show the physical structures of the models used in this study. Figures 2(a) and (b) show the 2D ($R-z$) distribution of gas density, and the equatorial-plane profiles of the density and velocity in the UCM model with our fiducial parameters (Table 1). For the latter, density and velocity are spatially averaged over $\theta - \pi/2 = [-0.03:+0.03]$. Mathematically, the density at the centrifugal radius diverges to infinity, but this averaged value reasonably represents the distribution around the equatorial plane. While the mass flux per streamline increases toward the equatorial plane, many streamlines intersect the equatorial plane around the centrifugal radius. The gas density thus has a sharp peak at the centrifugal radius in the equatorial plane. Due to the outflow cavity ($\theta_0 < 45$ deg), there is no gas inside the radius at which the streamline of $\theta_0 = 45°$ intersects the equatorial plane ($\approx$100 au). The density structure is consistent with the numerical simulations (e.g., Machida 2014) and recent observations (Lee et al. 2021) showing that the outflow is ejected from the inner part of the disk while the envelope gas accretes in the outer region.

Figure 2 (c) shows the temperature distribution, which is derived from the radiative transfer calculation. The temperature distribution can also be divided into two regions: the cavity region and the envelope region. The cavity region, which is assumed to be filled with tenuous (i.e., optically thin) gas in the radiation transfer calculation, has a spherically symmetric temperature distribution. The radiation entering the envelope region is absorbed and re-emitted at longer wavelengths, which determines the thermal structure in the envelope. A part of the absorbed energy is transported toward the equatorial plane, as in the two-layer temperature model of protoplanetary disks (e.g., Chiang & Goldreich 1997). The gas is thus cooler in the envelope than in the cavity region.

Since the SB model has been used to analyze the envelope structure of protostars (e.g., L1527) in previous studies (Sakai et al. 2014b), it is useful to compare our fiducial model based on the UCM model with the SB model. The SB model often assumes the emissivity distribution is proportional to $r^{-1.5}$ (e.g., Oya et al. 2022). We thus assume a uniform temperature distribution and a density distribution proportional to $r^{-1.5}$ for the SB model. Also, we set the cavity angle to be 80° (Oya et al. 2014) to restrict the flow near the equatorial plane (see Oya et al. 2022). The velocity field is given as follows:

$$v_r(r, \theta) = \sqrt{\frac{2GM}{r} - \frac{j^2}{r^2}}, \quad (14)$$

$$v_\theta(r, \theta) = 0, \quad (15)$$

$$v_\phi(r, \theta) = \frac{j}{r}, \quad (16)$$

where the angular momentum $j$ is assumed to be uniform over the envelope, and the total mechanical energy is conserved. This is derived from the angular momentum and kinetic energy conservation of a gas particle that has no energy and orbits in the midplane. The SB model cannot be defined inside $R_{CB}$, which corresponds to the pericenter of elliptically orbiting

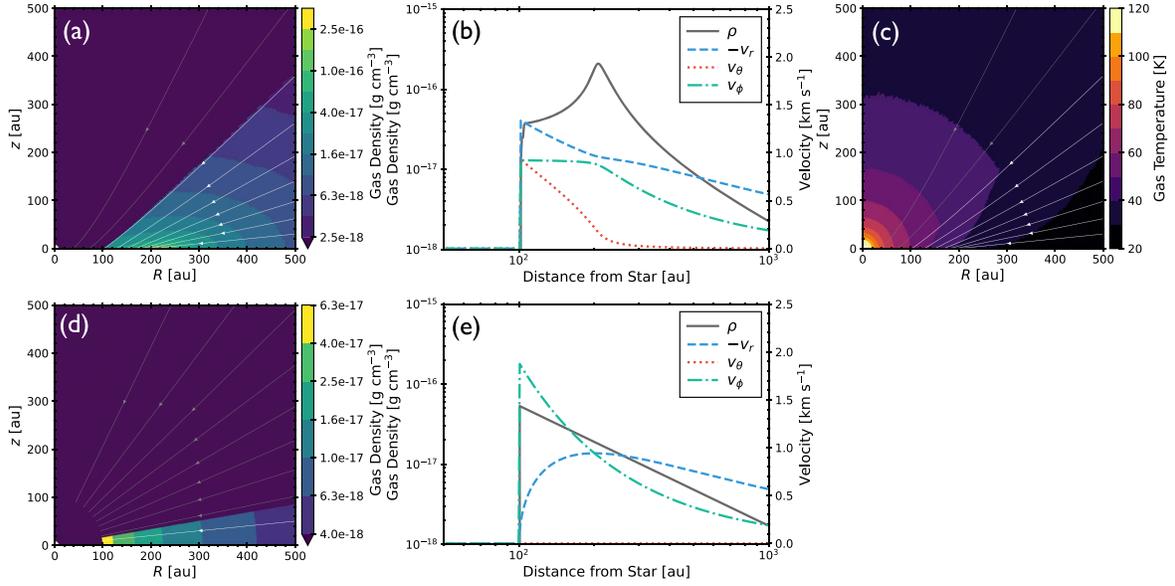

**Figure 2.** The upper panels show (a) the 2D ($R-z$) distributions of gas density, (b) the density and velocity distributions in the equatorial plane, and (c) the 2D temperature distribution of the UCM model with our fiducial parameters. The white lines in (a) and (c) depict the streamlines of gas for every 0.1 of $\cos(\theta_0)$, while the streamlines in the cavity are shown by gray lines. The mass flux per streamline is constant, and thus where the streamlines are close together indicates areas of high density. The lower panels show (d) the 2D distribution of gas density and (e) the density and velocity distributions in the equatorial plane in the SB model (see text). We do not display the temperature distribution of the SB model because it is uniform. In the equatorial plane distributions (b) and (e), the density and velocity are spatially averaged over $\theta - \pi/2 = [-0.03:+0.03]$.

particles. The radius of the centrifugal barrier is given as

$$R_{\rm CB} = \frac{j^2}{2GM}, \quad (17)$$

which is half of the centrifugal radius $R_{\rm CR}$.

Mathematically, Equation (7) has two kinds of solutions in the equatorial plane ($\theta = \pi/2$): the flow along the equatorial plane $\theta_0 = \pi/2$ and the accretion flow from the upper (lower) hemisphere $\theta_0 \neq \pi/2$. The latter solution is similar to the former near the equatorial plane at a radius much larger than the centrifugal radius, but the two solutions split up near or inside the centrifugal radius. The UCM model represents the accretion flow of $\theta_0 \neq \pi/2$, while the velocity distribution of the SB model (Equations (14)–(16)) is equivalent to the solution in the equatorial plane ($\theta = \pi/2$). The SB model can thus be regarded as a ballistic trajectory model limited to the flow in the equatorial plane for $r \gtrsim R_{\rm CR}$.

In the regions near and within the centrifugal radius, however, the SB model is quite different from the flow given by Equation (7) at $\theta = \pi/2$. In the SB model, the gas flow enters inside the centrifugal radius, while the flow of the UCM model (in the limit of $\theta = \pi/2$) does not. In other words, the SB model ignores the collision with the accretion flow from a high latitude or with the outer edge of the Kepler rotation disk. We also note that the gas density is significantly lower in the SB model than in the UCM model, since the SB model ignores the flow from the region of $\theta \neq \pi/2$ to the equatorial plane.

The distributions of gas density and velocity in the SB model are shown in Figures 2(d) and (e). The stellar mass is set to be the same as that in our fiducial model, while the angular momentum of the infalling gas is set to be the same as $j_{\rm mid}$. As discussed above, near the equatorial plane at $r \gg R_{\rm CR}$, the density and velocity fields in the SB model are very similar to those in the UCM model. At smaller radii, however, both density and velocity are quite different in the two models, since the flow from the vertical direction becomes dominant in the UCM model.

### 3.2. Position–Velocity Diagrams

We perform synthetic observations of the UCM models described in the previous subsection to understand how observational features are shaped. The left panel in Figure 3 shows the distribution of the integrated intensity $I$ normalized by its maximum value $I_{\rm max}$ for the UCM model with our fiducial parameters. The distribution of the integrated intensity is similar to a ring viewed edge on, because the gas is concentrated at the centrifugal radius (Figure 2 (a)). It looks similar to the integrated intensity map of c-$C_3H_2$ toward L1527 (Sakai et al. 2014b).

The right panel in Figure 3 shows the P–V diagram of the UCM model along the equatorial plane. The redshifted (blueshifted) components due to the rotation motion dominate in the P–V diagram for $x > 0$ ($x < 0$), while the blueshifted (redshifted) components originating from the infalling motion are also seen for $x > 0$ ($x < 0$), since we assume an inclination angle of 95°. This P–V diagram is also qualitatively similar to the P–V diagram of the c-$C_3H_2$ emission toward L1527. Especially in the first and third quadrants, as the offset decreases, the maximum velocity in the line of sight increases, which is called a spin-up feature. The velocity reaches the peak value at $\approx 120$ au, and then decreases inwards. In previous work (e.g., Sakai et al. 2014b), the position of the termination of the spin-up feature has been interpreted as the centrifugal barrier assuming the SB model (Sakai et al. 2014b). The UCM model, on the other hand, does not have a centrifugal barrier. Which physical structure or parameter then determines the location of the termination of the spin-up feature?

In order to better understand the characteristic feature of the P–V diagram of the UCM model, we map the line-of-sight velocity and the product of the density and temperature (i.e., $\rho T$) on the equatorial plane in Figure 4. The latter measures the brightness of the molecular emission (see Section 2.3) and is normalized by its maximum. The envelope gas rotating counterclockwise is observed from $y = -\infty$. For the outer region ($|x| \gtrsim 200$ au), the spin-up feature basically originates from the radial infall motion, but the rotation makes the velocity direction closer to the line-of-sight direction and thus increases the line-of-sight velocity. The location of the maximum line-of-sight velocity in the P–V diagram ($|x| \approx 120$ au; $|V| \approx 1.4$ km s$^{-1}$) corresponds to the region inside the centrifugal radius, where the velocity is due to the obliquely incoming accretion gas. Furthermore, the termination of the spin-up feature can be understood as a decrease in the emissivity and volume inside the centrifugal radius. Although there is gas with higher velocities inside $|x| \sim 120$ au, the emissivity and volume of the gas is small. In addition, there is a cavity inside $|x| = 100$ au.

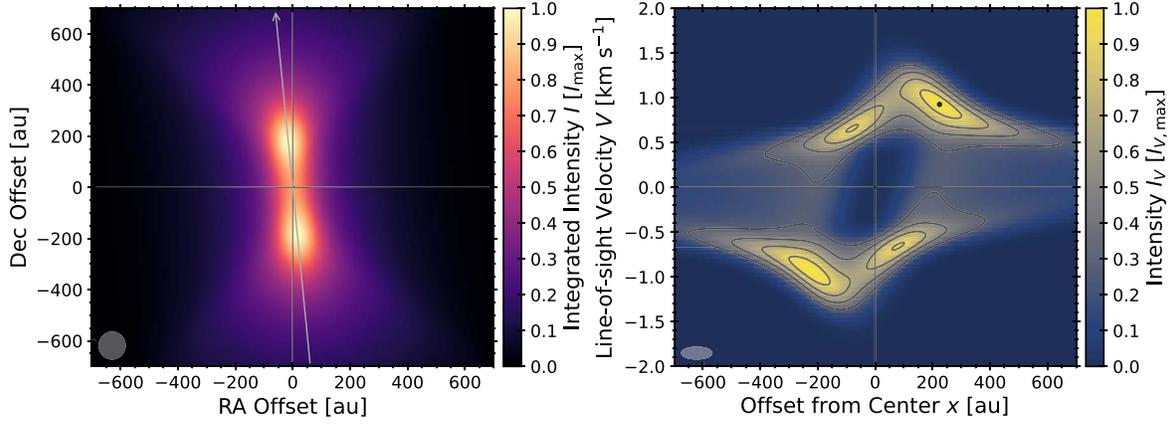

**Figure 3.** Left: integrated intensity (moment 0) map of our fiducial UCM model with an inclination of 95°. The intensity is normalized by its maximum value. The ellipse at the bottom left corner shows the convolution size. Right: P–V diagram along the gray arrow in the left panel, where the intensity is also normalized by its maximum value in the diagram. The black contours show 30%, 50%, 70%, and 90% of the maximum intensity.

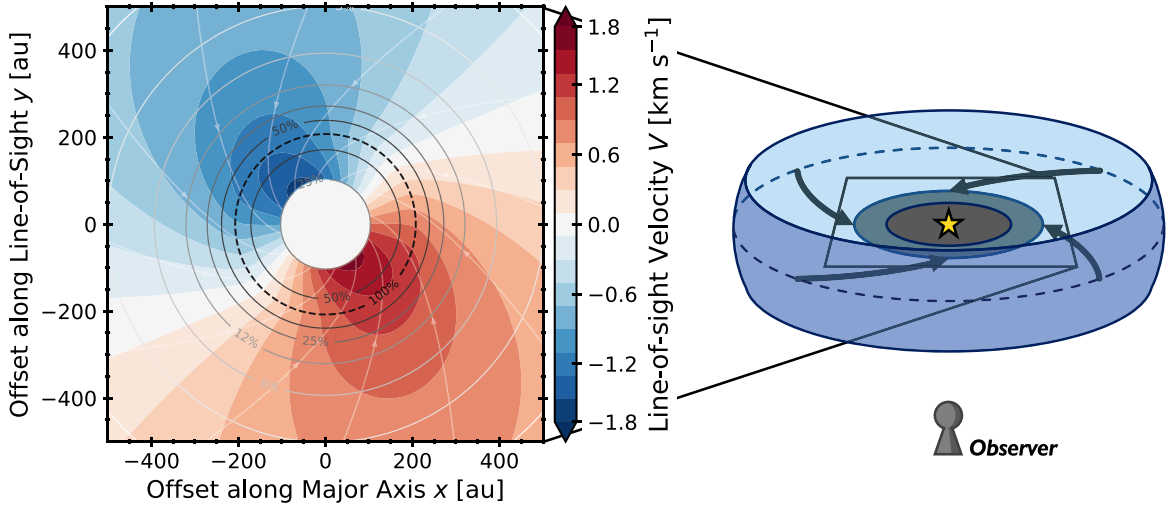

**Figure 4.** Distribution of the line-of-sight velocity on the equatorial plane (colors) for the UCM model. The contours are the product of density and temperature, to which emissivity is proportional, with levels decreasing by a factor of 2 from the maximum. The maximum value is shown by the dashed circle. The streamlines of the infalling rotating envelope are also plotted with white arrows.

While the SB model is characterized by the centrifugal barrier, the UCM model is characterized by the centrifugal radius, where the gas density has a sharp peak. Indeed, the moment 0 maps show an intensity peak at around the centrifugal radius. One may expect an intensity peak at the centrifugal radius in the P–V diagram, which could be used to estimate the central stellar mass. In the P–V diagram (Figure 3, right), the offset positions of the local intensity peaks are $x \approx 230$ au and $-80$ au for the positive line-of-sight velocity. Indeed, the former position may seem to be consistent with the position of the centrifugal radius $R_{\rm CR} \approx 200$ au. However, we find that the position on the equatorial plane that contributes most to the intensity peak is outward of the centrifugal radius.

Figure 5 shows the P–V diagrams for the same physical model as in Figure 3, but we limit the radius where the gas is present to inside and outside 250 au. We can see that the region outside the centrifugal radius contributes most to the intensity peak. As seen in Figure 4, the isovelocity contours happen to be almost parallel to the line of sight over a significant distance at $x = 230$ and $-80$ au. In other words, the position of the intensity peak in the diagram is determined by a geometrical effect.

In Figure 6, we compare the P–V diagram of the UCM model and the SB model (see Figures 2(d) and (e) for the physical structure of the SB model). While the P–V diagram of the SB model is almost the same as that of the UCM model at the offset $\gtrsim 300$ au, the SB model shows a higher velocity component near the center; along the 30% contour, the maximum line-of-sight velocities in the SB model and the UCM model are 1.53 and 1.36 km s$^{-1}$, respectively. In the SB model, all the gas kinetic energy is converted into the rotational energy at the centrifugal barrier. Thus, the SB model has a larger rotational velocity $v_\phi$ inside the centrifugal radius (see Figures 2(b) and (e)), resulting in a higher line-of-sight velocity.

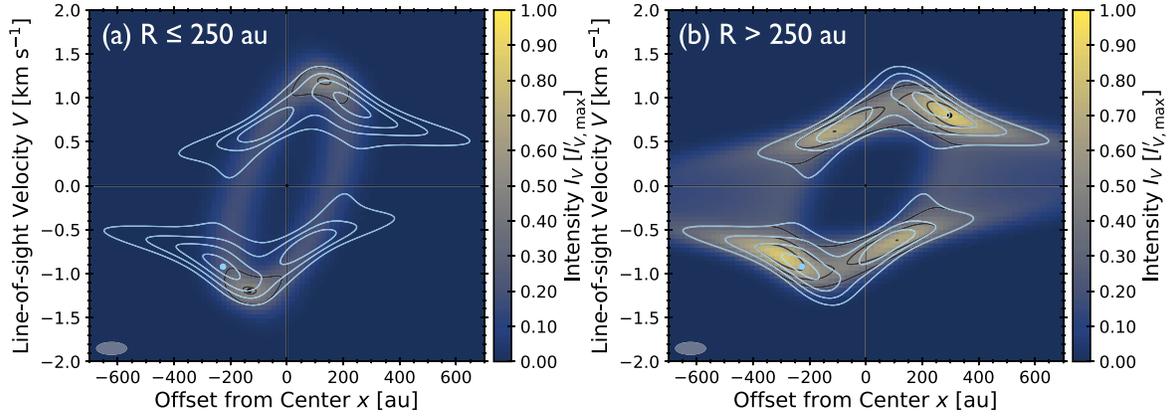

**Figure 5.** *P–V* diagrams of the UCM model for the same parameter set as in Figure 3, but with the region where gas is present restricted to (a) *R* < 250 au and (b) *R* > 250 au. This restriction is done after the calculation of the thermal structure. The contours show the intensity distribution of Figure 3. Both intensity distributions are normalized by the maximum intensity in the cube data used in Figure 3.

### 3.3. Parameter Dependencies

So far, we have chosen model parameters referring to L1527 as an example. Here, we investigate how the *P–V* diagram of the UCM model varies with input model parameters. We specifically pay attention to the maximum line-of-sight velocity and the position of the intensity peak (see also the mass estimation in Section 4.2).

Figures 7(a) and (b) show the *P–V* diagrams for cavity angles $\theta_{cav}$ of 0° and 80°, respectively, while the other parameters are set to be the same as in the fiducial model. At the zero-cavity angle, the envelope gas reaches the center on the equatorial plane. However, the density inside the centrifugal radius is still small, and thus the maximum line-of-sight velocity $V_{vmax}$ is not very different from that of the fiducial model. The offset position of the peak intensity does not change, either. Conversely, in the case of $\theta_{cav} = 80°$, $V_{vmax}$ decreases and $x_{vmax}$ increases. The shift of these values can be understood referring to Figure 4; with $\theta_{cav} = 80°$, there is no envelope gas inside the radius of 100 au in the equatorial plane.

Figures 7(c) and (d) show the *P–V* diagrams with a centrifugal radius of 100 and 300 au, respectively. $x_{vmax}$ and $x_{ipeak}$ show larger offsets for models with larger centrifugal radii. $x_{vmax}$ is 64 and 179 au in the models with $R_{CR} = 100$ and 300 au, respectively. In the model with $R_{CR} = 100$ au, $|x_{ipeak}|$ is 160 au, which is significantly larger than the centrifugal radius due to the beam dilution; the emission around the centrifugal radius is diluted by the observation beam (spatial convolution size) of ≈100 au.

Figures 7(e) and (f) show the *P–V* diagrams with $M = 0.1 M_\odot$ and $M = 0.3 M_\odot$, respectively. The centrifugal radius is the same as the fiducial model, and thus the specific angular momentum is larger for a larger mass (see Equation (17)). In other words, the model with a higher stellar mass corresponds to a more evolved stage, at which the envelope gas is accreting from a greater distance and with higher angular momentum. In the *P–V* diagram, the maximum line-of-sight velocity increases with the stellar mass, while the offset of the intensity peak does not significantly change. In the model with $M = 0.1 M_\odot$, the maximum line-of-sight velocity at 30% of the maximum intensity is 0.99 km s$^{-1}$, while it is 1.66 km s$^{-1}$ in the model with $0.3 M_\odot$.

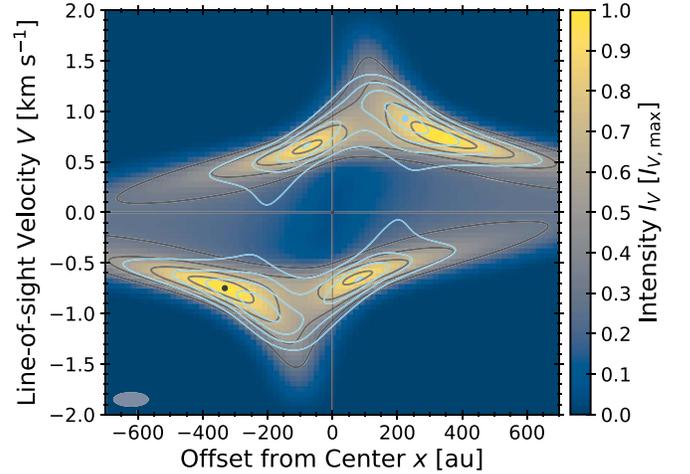

**Figure 6.** *P–V* diagram for the SB model with the fiducial parameter set (color map and black contours). The blue contours represent the *P–V* diagram for the UCM model (Figure 3). Both the contour levels are at 30%, 50%, 70%, and 90% of the maximum intensity, and the point shows the maximum intensity.

We next examine the effect of the inclination angle on the *P–V* diagram. When the inclination angle deviates from ∼90° (edge-on view), the diagram can be more complex because the velocity of the gas outside the equatorial plane contributes to the line-of-sight velocity. The diagram can also differ significantly depending on whether or not the line of sight intersects the outflow cavity. The left panels of Figure 8 show the moment 0 maps of our fiducial UCM model with inclination angles of 60° (upper panels) and 30° (lower panels), where the cavity angle is 45°. The right panels of Figure 8 show the *P–V* diagram along the major axis of the moment 0 map. In the case of $i = 60°$, the intensity peak of the moment 0 map is around the centrifugal radius (160 au; upper left panel in Figure 8). As in the fiducial model, this brightness distribution reflects the density distribution in the UCM model; the gas density reaches the peak value at the centrifugal radius and decreases toward the center at the inner radius (Figure 2). Emission at the central region is fainter than in the fiducial model but is not zero, since we have infalling envelope gas in the line of sight.

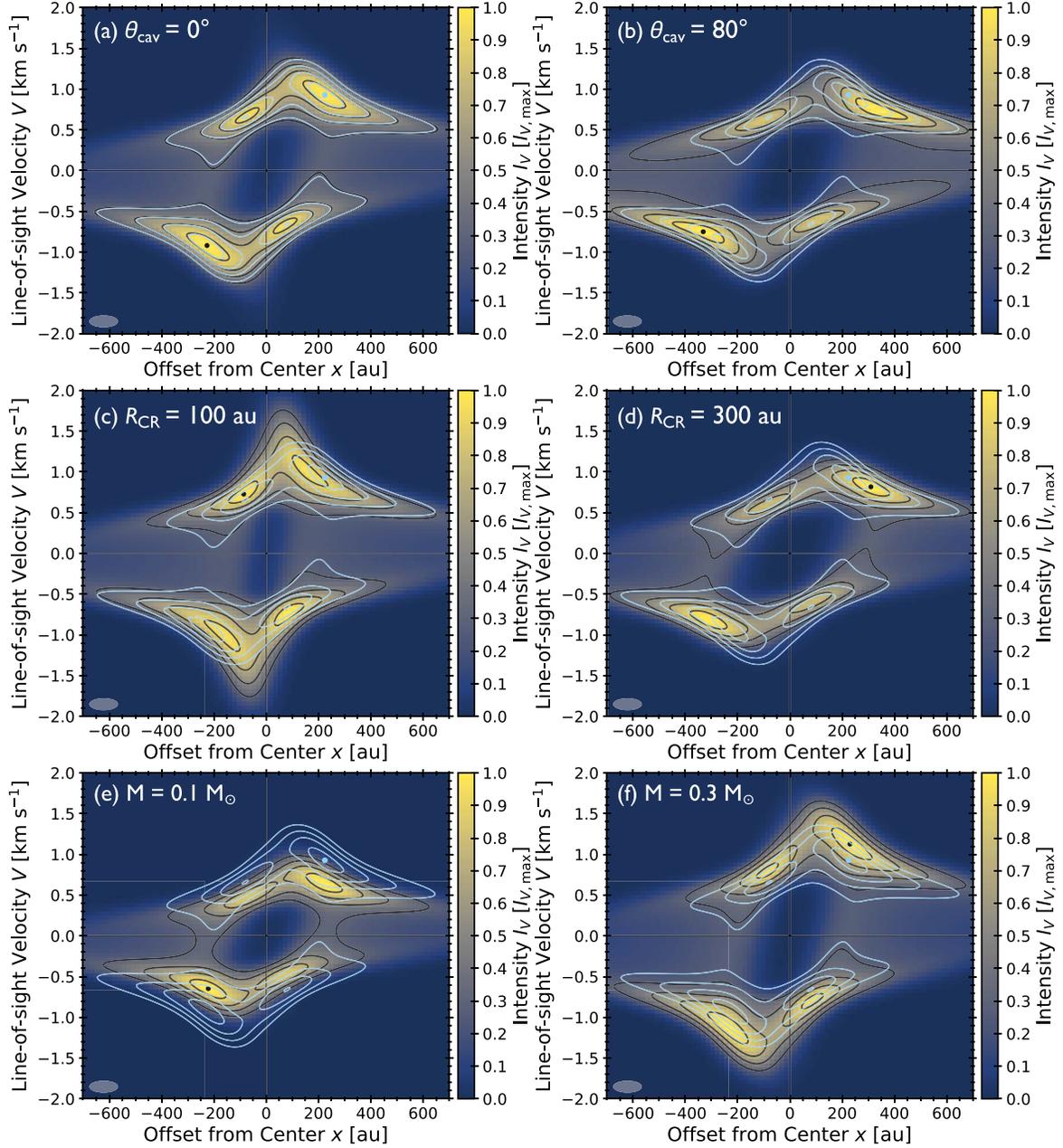

**Figure 7.** P–V diagrams (color map and black contours) of the UCM model with the same parameters as the fiducial model but (a) with a cavity angle of 0° and (b) 80°, (c) a centrifugal radius of $R_{CR} = 100$ au and (d) $R_{CR} = 300$ au, and (e) a protostellar mass of $M = 0.1\ M_\odot$ and (f) $M = 0.3\ M_\odot$. The P–V diagram of the fiducial model is overlaid as blue contours. The contour lines are the same as those in Figure 6.

The P–V diagram (upper right panel in Figure 8) is similar to that of the fiducial model except that the emission at the offset $|x| \lesssim 100$ au appears faint. The offset of the maximum line-of-sight velocity $|x_{vmax}|$ is ≈153 au and is shifted outward from the fiducial value. The moment 0 map with $i = 30°$ (lower right panel in Figure 8) shows a clear central cavity, since there is no envelope gas along the line of sight. The cavity is also apparent in the P–V diagram. In this case, the maximum line-of-sight velocity is mainly due to the obliquely accreting envelope rather than the flow on the equatorial plane. It should be noted that the above features also depend on the relative size of $R_{CR}$ and the observation beam.

### 3.4. Comparison of the Model Cube Data with the Observation of L1527

In this section, we examine how well the UCM model reproduces the observations by comparing the model cube data with the c-$C_3H_2$ observation of L1527 (Sakai et al. 2014b). The cube data contain more information than the P–V diagram and are therefore expected to give better constraints on the model parameters. We also obtain the best-fit model parameters for L1527 from the comparison. We further perform similar analysis with the SB model to investigate how different the obtained parameters are between the UCM and SB models. We vary the stellar mass and centrifugal radius in the synthetic observations for the models to obtain the best-fit parameters, while the cavity angle and inclination are fixed at 45° and 95°, respectively (Tobin et al. 2008; Sakai et al. 2014a, 2014b).

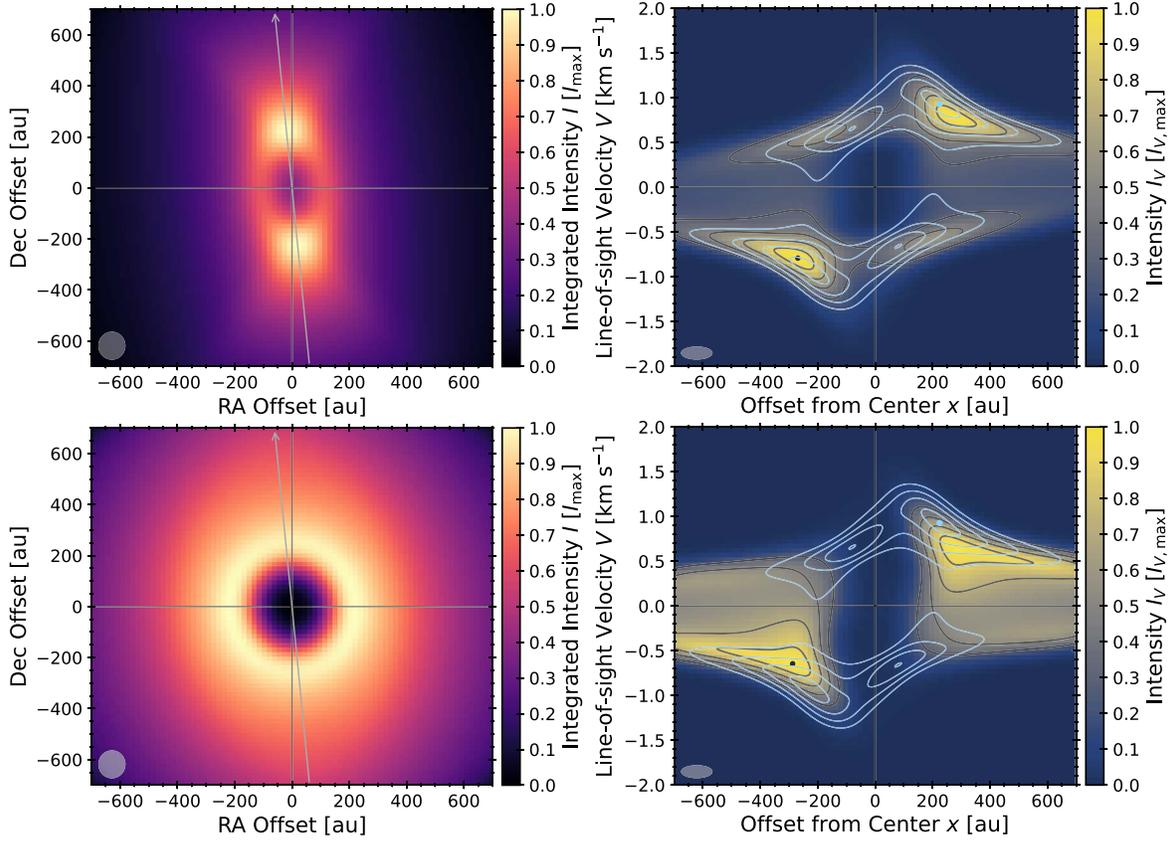

**Figure 8.** The moment 0 maps and the *P–V* diagrams of the fiducial UCM model but with inclination angles of 60° (upper) and 30° (lower). The *P–V* diagram of the fiducial model is overlaid as white contours. The contour lines are the same as those in Figure 6.

We compute the correlation coefficient known as the zero-mean normalized cross correlation, commonly used for template matching. Let us consider the correlation of the two cube datasets, $A(i, j, k)$ and $B(i, j, k)$ with size $(L, M, N)$. The grid $(i, j, k)$ indicates $(X_i, Y_j, V_k)$, and $X$ and $Y$ are the offsets along R.A. and decl., respectively. The correlation $R_{\rm col}$ is defined as

$$R_{\rm col} = \frac{\sum_{i=1}^{L}\sum_{j=1}^{M}\sum_{k=1}^{N}(A(i,j,k)-\bar{A})(B(i,j,k)-\bar{B})}{\sqrt{\sum_{i=1}^{L}\sum_{j=1}^{M}\sum_{k=1}^{N}(A(i,j,k)-\bar{A})^2 \cdot \sum_{i=1}^{L}\sum_{j=1}^{M}\sum_{k=1}^{N}(B(i,j,k)-\bar{B})^2}},$$
(18)

where the overbar denotes the average over all pixels. For example,

$$\bar{A} = \frac{\sum_{i=1}^{L}\sum_{j=1}^{M}\sum_{k=1}^{N}A(i,j,k)}{L \cdot M \cdot N}.$$
(19)

This correlation, ranging from −1 to 1, is a measure of the similarity of intensity *patterns* independent of the absolute value of the intensity: a value of 1 indicates identical datasets, 0 indicates no correlation, and a negative value indicates anticorrelation. We consider the data in the R.A. offset range of [−200, 200] au, the decl. offset range of [−700, 700] au, and the line-of-sight velocity range of [−3, 3] km s$^{-1}$, which cover the dominant emission region from the envelope. To mitigate the effect of noise in the observational data, we set a threshold value at 30% of the maximum intensity, and any data points below this threshold in both observational and model cube data are set to 0 for the analysis.

The correlation coefficient between the observational data and the UCM model reaches the maximum value of 0.47 with $M = 0.154\,M_\odot$ and $R_{\rm CR} = 290$ au, which we hereafter refer to as the best-fit parameters. Considering the number of independent data points (≈3000), the correlation coefficient of 0.47 indicates a good agreement between the model and the observational data. The variation of the correlation coefficient in the model parameter space $(M, R_{\rm CR})$ is gentle, even around the best-fit parameters, implying that a relatively wide range of $M$ and $R_{\rm CR}$ are in reasonable agreement with the observation.

The *P–V* diagram of the UCM model with the best-fit parameters along the major axis of the moment 0 map is compared with the observational data in the left panels of Figure 9. Although the peak velocity in the inner regions deviates from that of the observation, the overall features agree reasonably well with the observation. Since the data points showing high velocities are limited compared to those showing low velocities, the latter is weighted more in the correlation calculation.

For comparison, we also perform similar calculations with the SB model. Here, we assume a cavity angle of 80° to be consistent with the model used in the previous study (Oya et al. 2015), which compared the SB model with the observation using *P–V* diagrams along various position angles. As in the UCM model, the centrifugal radius in the SB model is defined as $R_{\rm CR} = j^2/GM$. This interpretation allows us to compare the results of the two models under the fixed angular momentum velocity. The maximum correlation coefficient is 0.49. In the right panels of Figure 9, we also show the P–V diagrams at three different cutting angles with best-fit parameters. Since the correlation coefficients of the best-fit models for the UCM model and the SB model are similar, both models reproduce the observation of L1527. In other words, we cannot conclude that either model is particularly superior. The best-fit parameters are $M = 0.158\,M_\odot$ and $R_{CR} = 248$ au ($R_{CB} = 124$ au), which are consistent with the parameters obtained from Sakai et al. (2014b) and Oya et al. (2015). While the best-fit centrifugal radius is larger in the SB model, the parameter ranges where the correlation coefficient is ⩾ 80% of the maximum value in the UCM model overlap with that of the SB model.

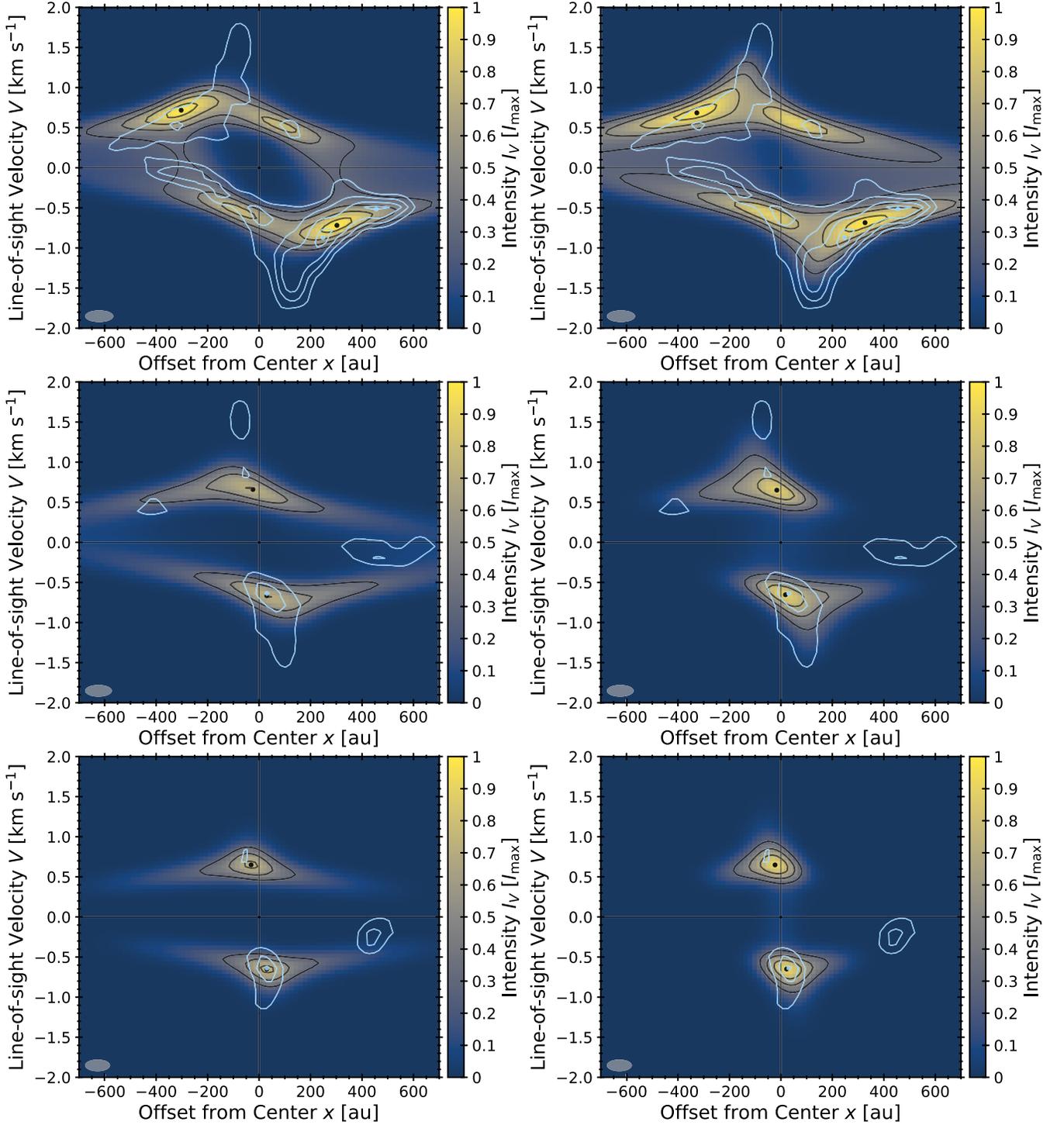

**Figure 9.** P–V diagrams obtained from the UCM model (left panels) and SB model (right panels), with color map and black contours, compared with those obtained from observations of c-$C_3H_2$ (Sakai et al. 2014b, blue contours) for the cutting angle of the P–V diagram of 0° (upper left), 30° (upper right), and 60° (lower) from the major axis (P.A. 5°, north to south). The levels of the contours are 30%, 50%, 70%, and 90% of the maximum intensity in each of the cube data. The points show the maximum intensity of each of the P–V diagrams.

One may notice that our synthetic P–V diagram does not show negative intensity around the center (low velocity and small offset), while it appears in the data of Sakai et al. (2014b). The negative intensity is due to the line absorption by the foreground cloud gas, which is not considered in our synthetic observation. Although this difference may in principle affect the correlation coefficient, the region of negative intensity is limited and would not change the relative value of the correlation coefficient among the models.

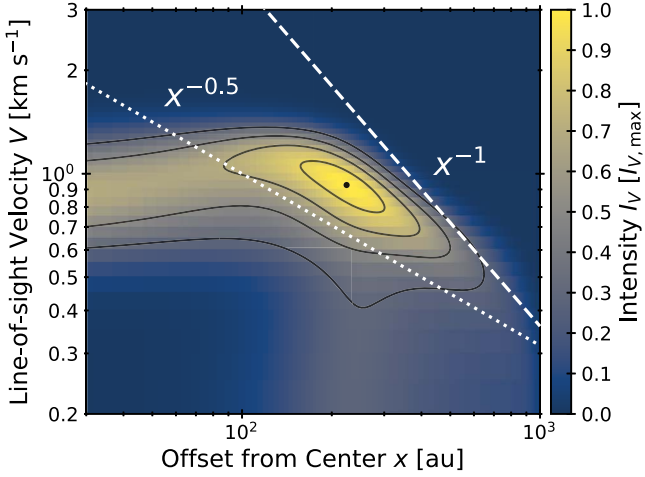

**Figure 10.** Same *P–V* diagram as Figure 3, but with logarithmic scales both for the offset position $x$ and line-of-sight velocity $V$ for the first quadrant ($x > 0$ and $V > 0$). Contour levels are the same as Figure 3. White lines show the power-law indices of $-0.5$ and $-1.0$ as a guide, in which the offsets are arbitrary.

## 4. Discussion

### 4.1. Slope of the Spin-up Feature of the Rotating Infalling Envelope

In this section, we analyze the envelope spin-up feature of the UCM model in detail. This is of particular interest because the feature is often used to distinguish between the disk and envelope. In a rotationally supported disk (i.e., Keplerian disk), the rotation velocity should be proportional to $R^{-0.5}$, while it is proportional to $R^{-1}$ if the gas falls with a constant specific angular momentum (e.g., Terebey et al. 1984; Ohashi et al. 1997, 2014; Takahashi et al. 2016; Sai et al. 2022). In the analysis of the CO emission from both the envelope and disk, the *P–V* diagram is compared with these power laws. In the case of L1527, the power-law index changes from $-1$ to $-0.5$ at $\approx 75$ au, which Aso et al. (2017) regard as the Keplerian disk size.

In Figure 10, we compare the *P–V* diagram for the UCM model with power laws of $x^{-0.5}$ and $x^{-1}$. We find that the velocity profile along the intensity ridge is closer to $x^{-0.5}$ than to $x^{-1}$ beyond $R \approx 200$ au. This is consistent with the synthetic observation in Sai et al. (2022), where a slope of $\sim -0.5$ is reported for the UCM model.

To understand why a $x^{-0.5}$ profile is obtained, we analytically estimate the profile of the line-of-sight velocity along the ridge in the *P–V* diagram. Here, we define the ridge velocity as the line-of-sight velocity at the maximum intensity in the velocity direction at each offset position (e.g., Yen et al. 2013; Ohashi et al. 2014). We assume an edge-on disk and calculate the radiative transfer (Equation (13)) from $y = +\infty$ toward the observer ($y = -\infty$) at each offset $x$

(Figure 4). We assume the local line profile to be a $\delta$ function with an amplitude $A(y)$, which can be written as $A(y) = \kappa_\nu \rho B_\nu$, and then convert the frequency into the line-of-sight velocity:

$$-\frac{\partial I_V(y)}{\partial y} = A(y)\delta(V - V_g(y)), \quad (20)$$

where $V_g(y)$ is the line-of-sight velocity of gas at $y$. The integration of this equation from $y = +\infty$ to $-\infty$ gives the intensity,

$$I_{V,\mathrm{obs}} = \int_{-\infty}^{+\infty} A(y)\delta(V - V_g(y))\,dy$$
$$= \sum_{V=V_g(y_*)} \frac{A(y_*)}{\left|\frac{\partial V_g}{\partial y}\right|_{y=y_*}}, \quad (21)$$

where $y_*$ is the location of $V = V_g(y_*)$. The distribution of $V_g$ in our fiducial model is shown in Figure 4. In general, the outer regions have a larger volume and thus have a larger contribution to the line intensity. We assume that most of the intensity comes from the outer region and $A(y)$ is approximately expressed as a power-law function of $R = \sqrt{x^2 + y^2}$, $A \approx A_0 R^{-p}$, where $A_0$ and $p$ are constant. The observed intensity at $x$ and $V$ is given by

$$I_{V,\mathrm{obs}} \approx \frac{A_0(x^2 + y_*^2)^{-p/2}}{\left|\frac{\partial V_g}{\partial y}\right|_{y=y_*}}. \quad (22)$$

Thus, to obtain the ridge velocity, we determine the value of $V$ at which $I_{V,\mathrm{obs}}$ reaches its maximum value for a given $x$.

First, we consider a simple case where the envelope has only free-fall velocity without rotation. In this case, the line-of-sight velocity is given as

$$V_g = \sqrt{2\frac{GM}{r}}\,\frac{y}{r}. \quad (23)$$

This equation can be scaled by the centrifugal radius and a typical velocity $V_x$ at the offset $x$ as

$$\frac{V_g}{V_x} = \sqrt{2}\,\tilde{y}\left(1 + \tilde{y}^2\right)^{-3/4}, \quad (24)$$

where $\tilde{y} = y/x$ and $V_x = \sqrt{GM/x}$. Differentiating $V_g$ by $y$, we obtain

$$I_{V,\mathrm{obs}} = \frac{\sqrt{2}\,A_0}{V_x}\,\frac{x\left(1 + \tilde{y}_*^2\right)^{7/4 - p/2}}{|\tilde{y}_*^2 - 2|}. \quad (25)$$

This function diverges to infinity at $\tilde{y}_* = \sqrt{2}$, i.e., at $V/V_x = (16/27)^{1/4} \approx 0.88$. Although this discussion is based on an infinite resolution, the intensity at the divergent point would take a maximum value by considering the beam dilution in reality. The important point here is that the line-of-sight velocity at the maximum intensity is proportional only to $V_x$. Therefore, substituting it into Equation (24), we find that the line-of-sight velocity $V_{\mathrm{ridge}}$ at the intensity maximum on the equatorial plane is proportional to $x^{-0.5}$. When the envelope

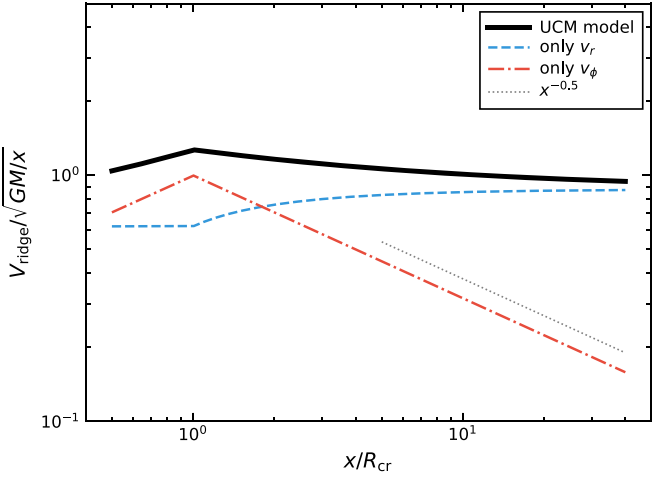

**Figure 11.** Profile of the line-of-sight velocity $V_{\rm ridge}$ at the location where the emitted energy takes the maximum on the line of sight at $x$. The offset $x$ from the center is normalized by the centrifugal radius $R_{\rm CR}$. Dashed, dashed–dotted, and solid lines show the cases of free fall, Keplerian rotation, and the UCM model, respectively. The dotted line is $x^{-0.5}$, for reference.

has only the rotational velocity where the gas retains its angular momentum,

$$V_g(x, y) = \frac{x v_\phi}{r} = \frac{j_{\rm mid}}{x(1 + \tilde{y}^2)}. \quad (26)$$

The same analysis leads to $V_{\rm ridge} \propto x^{-1}$.

In the UCM model, we expect $V_{\rm ridge} \propto x^{-0.5}$, since the radially infalling gas is the main contributor to the maximum intensity. To confirm this estimate, in Figure 11 we plot the line-of-sight velocity at which $\partial V_g / \partial y = 0$ for the UCM model, i.e., the velocity along the ridge in the P–V diagram. We see that $V_{\rm ridge}/\sqrt{GM/x}$ is mostly constant, i.e., $V_{\rm ridge}$ is proportional to $x^{-0.5}$. The blue dashed line shows the ridge velocity estimated only from the radial motion, while the red dotted line considers only the rotational velocity. In the outer region, the constant $V_{\rm ridge}$ is due to the radial velocity. At smaller offsets $x$, the contribution of $v_r$ decreases, while the rotational velocity increases. As a result, $V_{\rm ridge}$ remains nearly constant.

Although the above analyses explain $V_{\rm ridge} \propto x^{-0.5}$ in the UCM model, the $C^{18}O$-line observations (Ohashi et al. 2014; Aso et al. 2015, 2017) suggest that the line-of-sight velocity $V_{\rm ridge} \propto x^{-1}$ in the envelope of L1527. There are two possibilities to account for this contradiction. First, $C^{18}O$ emission is contaminated by the slower velocity component in the line of sight, since the critical density ($10^4$ cm$^{-3}$ for $J = 2 - 1$ line) is rather low. This may result in an apparently slower velocity being observed. The contamination may be mitigated by selecting observed molecular lines (Oya & Yamamoto 2020; Oya et al. 2022). Alternatively, this contradiction may imply that the radial gas velocity is significantly slower than the free-fall velocity, as Sai et al. (2022) have shown and Ohashi et al. (2014) and Chou et al. (2014) suggested. The infall motion of the envelope is more likely to be affected by, e.g., magnetic fields than the Keplerian motion of the disk, since the kinetic energy of the former is smaller than the latter. When the radial velocity is reduced, the rotational velocity is mainly determined by the line-of-sight velocity, leading to $V_{\rm ridge} \propto x^{-1}$.

### 4.2. Mass Estimation of the Central Star from the Specific Features in the P–V Diagram

Comparing observational data with a model through radiative transfer modeling (i.e., synthetic observations) is useful for interpreting the observational data and for deriving physical parameters of the target object. There are, however, various methods of comparison. In Section 3.4, we estimated the central stellar mass by calculating the correlation between the synthetic cube data with the observational data. In this evaluation, all pixels on the plane of the sky are weighted equally, as long as the intensity of each pixel is higher than a threshold value. Alternatively, we can compare a specific feature of the data.

Previous works by Sakai et al. (2014b) and Oya et al. (2014) focused on the termination point of the spin-up feature in the P–V diagram. In the line observation of ETMs toward L1527, Sakai et al. (2014b) showed that the maximum rotational velocity at the termination point is two times larger than the maximum radial velocity at the zero-offset position, which motivated them to invent and adopt the SB model (see also Oya et al. 2022). In this model (Equations (14)–(16)), the maximum line-of-sight velocity $V_{\rm vmax}$ and its position $x_{\rm vmax}$ correspond to the radius and velocity of the centrifugal barrier,

$$R_{\rm CB} = \frac{j_{\rm mid}^2}{2GM}, \quad (27)$$

$$v_{\rm CB} = \frac{2GM}{j_{\rm mid}}, \quad (28)$$

while the maximum velocity at the zero-offset position $V_{x=0}$ is the maximum radial velocity $\max(|v_r|)$. In Figure 6, we can indeed see that the ratio of $V_{\rm vmax}/V_{x=0}$ is relatively high in the SB model compared with the ratio in the UCM model, although it is smaller than 2, probably due to smoothing by the beam. By eliminating $j_{\rm mid}$ from Equations (27) and (28), $M$ can be derived from $R_{\rm CB}$ and $v_{\rm CB}$:

$$M_{\rm vmax} = \frac{1}{2G} x_{\rm vmax} \left(\frac{V_{\rm vmax}}{\sin(i)}\right)^2. \quad (29)$$

While the mass estimate based on the SB model is simple and thus relatively easy to apply to observational data, we examine here the systematic difference caused by assuming this model. We apply this method to the synthetic P–V diagram of the UCM model and compare the estimated stellar mass of the SB model with that of the UCM model. Specifically, we calculate the P–V diagram based on the UCM model with different stellar masses and centrifugal radii (Section 3.3). We use an intensity of 30% of the maximum value as the threshold signal. This threshold intensity (i.e., signal-to-noise threshold) is given as a comparable value to that of the observation of L1527 (Sakai et al. 2014b). Then we measure $V_{\rm vmax}$ and $x_{\rm vmax}$ to derive $M_{\rm vmax}$ from Equation (29). The opening angle of the cavity is set to 45°.

The left panels in Figure 12 present the ratio of $M_{\rm vmax}$ to the mass in the UCM model, as a function of the centrifugal radius and model mass. The upper panels are for the edge-on system, while the lower panels are for an inclination of 60°.[8] The ratio is smaller than unity, except for a limited parameter space in the lower panel. The mass is underestimated due to the difference in velocity structure between the UCM model and the SB model. In the UCM model, the maximum velocity is due to the gas accreting obliquely onto the midplane, while in the SB model, it is due to the high rotational velocity at the centrifugal barrier. Therefore, the maximum velocity in the UCM model is lower than that in the SB model for the same mass. It should also be noted that the mass estimate based on the maximum line-of-sight velocity depends on the threshold intensity, which is set to be 30% of the maximum value of intensity in the present study. Furthermore, since the oblique accretion flow inside the centrifugal radius is seen as the maximum velocity, this mass estimate based on $V_{vmax}$ could depend on the outflow cavity angle.

---
[8] That is, 30° deviation from edge on.

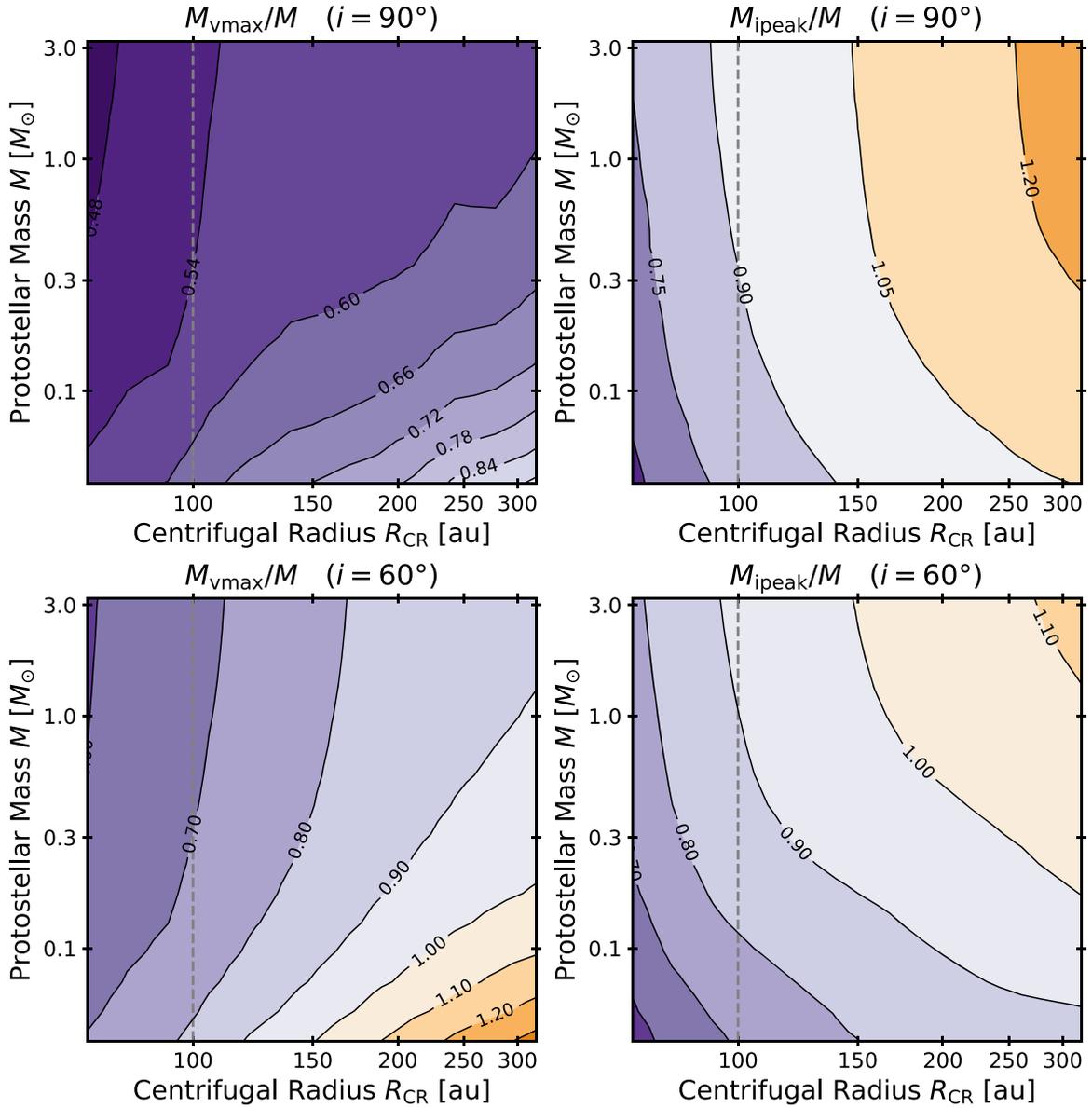

**Figure 12.** The ratio of the estimated stellar mass in the SB model to the mass in the UCM model is a function of centrifugal radius and stellar mass. In the left panels, the stellar mass is estimated from $v_{max}$ and $x_{vmax}$, assuming the centrifugal barrier (i.e., SB model). In the right panels, we assume that the peak intensity in the P–V diagram corresponds to the centrifugal radius. The upper panels are for edge-on geometry, while the lower panels are for an inclination angle of 60°.

We also estimate the stellar mass by using the offset position $x_{ip}$ and velocity $V_{ip}$ at the intensity peak in the P–V diagram. As we have seen in Section 3.2, these quantities are roughly related to the physical structure of the infalling rotating envelope around and outside the centrifugal radius due to the geometrical effect. As $V_{ip}$ scales with the typical velocity $V_{CR}$ and $x_{ip}$ scales with the centrifugal radius $R_{CR}$, the central mass is proportional to $x_{ip} V_{ip}^2 / G$. Here, we simply give the mass estimate as

$$M_{ip} = \frac{1}{G} x_{ip} V_{ip}^2. \quad (30)$$

While the linear scaling could be affected by the observational resolution, we examine here how good or bad this estimate is. The right panels in Figure 12 again show the estimated-to-model mass ratio in the UCM model. For both the edge-on system and a 60° inclination, the estimated mass is within ∼0.7–1.2 of the model mass in a large parameter space.

We should note that the actual observations are not as clean as the model observations. The protostellar envelope in reality has nonsymmetric distribution of mass and velocity, which affects the intensity distribution and thus the stellar mass estimate using the specific local features.

### 4.3. Free Fall or Slow Infall

One of the motivations of the present work is the discrepancy of the stellar masses estimated from Keplerian rotation and from the infalling envelope. For L1527, Aso et al. (2017) derived $M = 0.45\ M_\odot$ by fitting the P–V diagram of $C^{18}O$ with Keplerian velocity, while Sakai et al. (2014b) derived $M = 0.18\ M_\odot$, assuming the SB model for the P–V diagram of the ETM emission. We investigated how the stellar mass estimate obtained using the ETM depends on the assumed envelope model.

In Section 3.4, we estimate the stellar mass of L1527 by calculating the correlation between the cube data of the observation and the models. The best-fit mass is $0.154\ M_\odot$ for the UCM model and $0.158\ M_\odot$ for the SB model. Both values are much lower than $0.45\ M_\odot$, which is derived from the analysis of Keplerian rotation (Aso et al. 2017). We can also estimate the stellar mass of L1527 based on the maximum velocity (Equation (29)) and the intensity peak (Equation (30)) for the quadrant where the spin-up feature appears and the intensity is stronger. Adopting a threshold intensity of 30% of the maximum intensity, the maximum line-of-sight velocity is 1.81 km s$^{-1}$ at $|x_{\rm vmax}| = 97$ au, and thus the stellar mass is $0.179\ M_\odot$. The intensity peak, on the other hand, is at $|v_{\rm ipeak}| = 0.92$ km s$^{-1}$ and $|x_{\rm ipeak}| = 250$ au. Then, Equation (30) gives $0.238\ M_\odot$. To correct for a possible error in these estimates, we refer to the edge-on case in Figure 12, as L1527 is almost edge on. For the mass estimate using the maximum line-of-sight velocity, simply assuming $R_{\rm CR} \sim 2|x_{\rm vmax}|$ and $M \sim 0.2\ M_\odot$, $M_{\rm vmax}/M$ is $\sim 0.6$. For the mass estimate using the peak intensity, assuming $R_{\rm CR} \sim |x_{\rm ipeak}|$ and $M \sim 0.24\ M_\odot$, $M_{\rm ipeak}/M$ is $\sim 1.1$. Then, the stellar mass with the error correction is $\sim 0.22$–$0.30\ M_\odot$, which is still lower than $0.45\ M_\odot$. We may therefore conclude that the stellar mass estimated from the infalling envelope is lower than that from Keplerian rotation. This result suggests that the infall could be retarded compared with the UCM model or free fall, possibly due to the magnetohydrodynamic effect (e.g., Chou et al. 2014; Ohashi et al. 2014; Aso et al. 2017; Sai et al. 2022).

The retardation of the infall gas might be supported by the higher ratio of $V_{\rm vmax}$ to $V_{x=0}$ in the observations (see Section 4.2; Sakai et al. 2014b). Whereas the SB model shows the ratio ∼2, the UCM model shows a lower ratio (e.g., see Figure 3), which seems to be insensitive to the model parameters. However, if the infall gas is retarded enough, the ratio could be higher because $V_{x=0}$ is reduced, while $V_{\rm max}$ is determined by the rotation speed and is maintained. In addition, the retardation could be responsible for the spin-up feature with $V \propto x^{-1}$ (rather than $x^{-0.5}$) in the P–V diagram (see Section 4.1).

It should be noted, however, that the $C^{18}O$ line could be contaminated with the innermost envelope component considering the beam size of the observation. The line-of-sight velocity of the innermost region of the envelope could be larger than the Keplerian velocity. The contamination could result in the overestimation of the Keplerian velocity and thus of the stellar mass.

### 4.4. Initial Angular Momentum Distribution

One of the major differences between the UCM model and the SB model is in the initial distribution of angular momentum; the UCM model assumes a rigid rotation of the core, while the specific angular momentum is assumed to be constant (uniform) in the SB model. In reality, the distribution of angular momentum of the infalling gas could be more complex, which could be one of the reasons why asymmetric P–V diagrams (e.g., Murillo et al. 2013; Sakai et al. 2016; Oya et al. 2018; Maureira et al. 2020; Kido et al. 2023) and warped disks (e.g., Sakai et al. 2019; Sai et al. 2020; Yamato et al. 2023) are often seen in observational data.

In recent years, asymmetric inflow (i.e., streamers) have been observed toward some protostars (e.g., Yen et al. 2019; Pineda et al. 2020; Tychoniec et al. 2021; Thieme et al. 2022; Aso et al. 2023; Kido et al. 2023). Non-axisymmetric streamers around the centrifugal radius can influence the intensity distribution in the cube data and the P–V diagram. While the shape of the streamers can be used to constrain the stellar mass, it requires the reconstruction of the 3D structure of the streamer.

On the theoretical side, Misugi et al. (2019) found that the observed angular momentum distribution among cores can be understood by the fragmentation of filaments with velocity fluctuations of 1D Kolmogorov spectra. The angular momentum distribution within the cores then depends on the balance between the dissipation of the turbulence within the core and inflow of filament gas to the core. A comparison of the observational data with such sophisticated core models remains to be investigated in future work (see Hanawa et al. 2022 for the hydrodynamic accretion of compact cloud clumps).

### 5. Summary

We performed synthetic observations of the UCM model to understand the relation between the physical structures of the infalling envelope around a protostar and the observational features probed by molecular lines. As an example, we constructed a UCM model adopting the physical parameters of L1527, and synthesized the cube data, integrated intensity map, and the P–V diagram along the direction perpendicular to the rotational axis. The integrated intensity map is in reasonable agreement with the observational data; it is similar to a ring viewed edge on since the gas is concentrated around the centrifugal radius. The P–V diagram is also similar to the observational data; it has a local intensity peak in each quadrant of the diagram and shows a spin-up feature. The spin-up feature is due to the increase of the infalling velocity toward the center and is terminated around the centrifugal radius as the volume and mass of the infalling gas decrease.

We compared the formulation, physical structure, and synthetic P–V diagrams of the UCM model and the SB model. We also performed synthetic observations of the UCM model with various parameters of centrifugal radius, central stellar mass, outflow cavity angle, and inclination angle to investigate how the P–V diagram changes with these parameters.

There are multiple ways to compare synthetic data with observational data. We first calculated the correlation coefficient, known as the zero-mean normalized cross correlation, between the observational data of L1527 and the UCM model by varying the stellar mass and $R_{\rm CR}$ (i.e., the angular momentum). A good correlation is obtained with the best-fit parameters of $M = 0.154\ M_\odot$ and $R_{\rm CR} = 290$ au. A similarly good correlation between the observational data and the SB model is obtained with $M = 0.158\ M_\odot$ and $R_{\rm CR} = 248$ au. While the correlation reflects the overall similarity between the cube datasets, we can alternatively compare local specific features. In previous studies, the stellar mass is often estimated from the position of the maximum line-of-sight velocity in the $P$–$V$ diagram assuming the SB model (i.e., centrifugal barrier). We have applied this method to the synthetic $P$–$V$ diagram of the UCM model to evaluate the systematic uncertainty. The stellar mass tends to be underestimated since the gas flows obliquely onto the equatorial plane in the UCM model rather than reaching the centrifugal barrier (i.e., perihelion), as assumed in the SB model. We also examined if we could estimate the stellar mass from the position of the maximum intensity in the $P$–$V$ diagram, taking advantage of the fact that the peak position is around the centrifugal radius. The ratio of the estimated mass and the actual mass in the model is $\sim 0.7$–$1.3$.

In the case of L1527, the stellar mass values estimated using the multiple methods discussed above are all lower than the value derived from the Keplerian analysis of the disk seen in $C^{18}O$. This may indicate that the gas infall motion in the envelope is retarded by, e.g., magnetic fields.

We analytically showed that the spin-up feature of the $P$–$V$ diagram is close to $V \propto x^{-0.5}$ in the UCM model, which is counterintuitive, since $V_{\rm rot}$ should be $\propto x^{-1}$ for infalling gas with a constant angular momentum. Indeed, for the observation of protostellar cores, the power of the velocity, $V \propto x^p$, is used to probe the transition between the infalling envelope ($p = -1$) and Keplerian disk ($p = -0.5$). The trick is that the line-of-sight velocity is dominated by infall rather than rotation in the UCM model. If the infall is retarded, rotational velocity should dominate so that $V$ is proportional to $x^{-1}$.

## Acknowledgments

The authors are grateful to Masahiro Machida, Shingo Hirano, Jinshi Sai, and Nagayoshi Ohashi for fruitful discussions. The authors also thank Anton Feeney-Johansson for carefully proofreading the draft. This work is supported by the Japan Society for the Promotion of Science KAKENHI grant Nos. JP18H05222, JP20H05844, JP20H05845, JP20H05847, JP21J00086, 22KJ0155, and 22K14081. We also acknowledge support from the RIKEN pioneering project, Evolution of Matter in the Universe.

## ORCID iDs

Shoji Mori https://orcid.org/0000-0002-7002-939X
Yuri Aikawa https://orcid.org/0000-0003-3283-6884
Yoko Oya https://orcid.org/0000-0002-0197-8751
Satoshi Yamamoto https://orcid.org/0000-0002-9865-0970
Nami Sakai https://orcid.org/0000-0002-3297-4497